\documentclass[10pt,letterpaper]{article}
\usepackage[top=0.85in,left=2.75in,footskip=0.75in]{geometry}

\usepackage{amsmath,amssymb}
\usepackage{changepage}
\usepackage[utf8]{inputenc}
\usepackage{textcomp,marvosym}
\usepackage{cite}
\usepackage{nameref,hyperref}
\usepackage{microtype}
\DisableLigatures[f]{encoding = *, family = * }
\usepackage[table]{xcolor}
\usepackage{array}
\usepackage{graphicx}
\usepackage{enumitem} 

\usepackage{braket}

\newcolumntype{+}{!{\vrule width 2pt}}

\newlength\savedwidth

\raggedright
\usepackage[aboveskip=1pt,labelfont=bf,labelsep=period,justification=raggedright,singlelinecheck=off]{caption}

\makeatletter
\renewcommand{\@biblabel}[1]{\quad#1.}
\makeatother

\usepackage{lastpage,fancyhdr,graphicx}
\usepackage{epstopdf}
\fancyheadoffset[L]{2.25in}
\fancyfootoffset[L]{2.25in}

\begin{document}
\vspace*{0.2in}

\begin{flushleft}
{\Large
\textbf\newline{Quantum annealing in spin-boson model: from a perturbative to a ultrastrong mediated coupling} 
}
\newline
\\
Manuel Pino*, Juan Jos\'e Garc\'ia-Ripoll\\
\bigskip
Institute of Fundamental Physics IFF-CSIC, Calle Serrano 113b, Madrid 28006, Spain\\

\bigskip
*mpg@iff.csic.es

\end{flushleft}
\section*{Abstract}
We study a quantum annealer where bosons mediate the Ising-type interactions between qubits. We compare the efficiency of ground state preparation for direct and  mediated couplings, for which Ising and spin-boson Hamiltonian are employed respectively. This comparison is done numerically for a small frustrated antiferromagnet, with a careful choice of the optimal adiabatic passage that reveals the features of the boson-mediated interactions. Those features can be explained by taking into account what we called excited  solutions: states with the same spin correlations as the ground-state but with a larger bosonic occupancy. For similar frequencies of the bosons and qubits, the performance of quantum annealing depends on how excited solutions interchange population with local spin errors. We report an enhancement of quantum annealing thanks to this interchange under certain circumstances. 


\section{Introduction}\label{Sec:1}

Quantum information theory provides us with various alternative and equivalent models of quantum computation. In the adiabatic quantum computer (AQC) model, the problem is formulated as a Hamiltonian $H_p$ whose ground state encodes the solution. The ground state is prepared slowly or ”adiabatically”\ \cite{albash2018adiabatic}, modulating the dynamics and interaction of a tunable quantum system —typically a quantum simulator of a spin Hamiltonian—, from a model and initial state that are easy to prepare $H(0) = H_0$, down to the final Hamiltonian $H(T) = H_p$ , where we expect to find the system in the solution state. Different choices of problem Hamiltonians, initial states and adiabatic passages correspond to different ``algorithms". Adiabatic quantum computers can be as powerful as universal quantum computers based on quantum gates\ \cite{aharonov2008adiabatic}, provided the Hamiltonian $H$ is a general enough spin model. However, the experimental and theoretical study of AQC algorithms has largely focused on combinatorial optimization problems\ \cite{farhi2001quantum}. These are problems that are of practical interest, include families of NP-hard problems and, most important, map to Ising models $H_p=\sum_{ij}\sigma^x_i\sigma^x_j$ that have been implemented using superconducting qubits\ \cite{johnson2011quantum} or trapped ions\ \cite{friedenauer2008simulating,kim2010quantum,britton2012engineered}. In this reduced framework of combinatorial problems, existing AQC devices behave very much like Monte Carlo Quantum Annealing methods for preparing ground states of spin Hamiltonians\ \cite{boixo2014evidence}. It seems that the performance of the adiabatic quantum algorithm is linked to the efficiency of quantum tunneling and quantum fluctuations to overcome energy barriers and reach the true ground state of the problem\ \cite{denchev2016computational}. Any advantage of physical devices over other stochastic simulations must therefore come from a more productive use of those fluctuations.

In this work we study the role of quantum fluctuations in adiabatic passages of Ising-type systems where the interactions are mediated by bosonic particles, such as the phonons of an ion trap\ \cite{porras2004effective} or superconducting squids coupling flux qubits\ \cite{harris2010experimental}. Such devices exhibit two sources of quantum fluctuations: the transverse magnetic fields that appear in $H_0=\sum_i\sigma^z_i,$ and the quantum fluctuations of the bosons that mediate the interaction. The last type of fluctuations are usually neglected in the (slow) perturbative coupling limit, where the interaction strength between spins and bosons $g$ exceeds the detuning between the bosonic excitation energy $\omega$ and the spin gap $\omega_0$ --i.e. $g\gg \delta = |\omega-\omega_0|$--. This is not the only possible operating point, as one may get the same effective spin-models in the ultrastrong coupling regime\ \cite{kurcz2014hybrid,schiro2012phase}, where couplings $g\simeq \delta$ and quantum fluctuations become an integral part of a much faster dynamics. The question we address in this work is whether bosonic fluctuations improve or damage the final outcome of the quantum adiabatic algorithm.

Specifically, we will present general conclusions about how bosonic fluctuations along a passage affect the performance of quantum annealers. Our intuition is that the excess of energy due to local errors, i.e. spin flips, can be transferred to the bosons, correcting the error. Similar questions have been addressed in the context of Landau-Zener transition of a single two-level system coupled to several bosonic modes\ \cite{wubs2006gauging,arceci2017dissipative}. These works indicate that adiabaticity improve or worsen depending on the spin-boson coupling. A work by Tian\ \cite{tian2018universal} also reports a speedup in a quantum annealer with pre-existing interactions that couples to a common bosonic mode\ \cite{tian2018universal}, due to an enhancement of the minimum gap by the bosonic mode. However, if we remove the direct spin-spin interactions, it seems that the common bosonic degrees of freedom can lead to a worsening of Landau-Zener passages, according to\ \cite{altland2008many}. A careful and detailed study of a realistic model is therefore required.

The structure of the paper is as follows. In next section, we explain why and how mediated interactions naturally appear in the implementation of a generic adiabatic quantum computer. We will show that the spin-boson Hamiltonian can be used to simulate a spin Hamiltonian where interactions are mediated by the bosons. In particular, the spin-boson coupling can be engineered to create a ground-state that encodes the solution of hard optimization problems. Section\ \ref{Sec:3} contains the main results of this work. We present the adiabatic theorem and how it imposes conditions for the AQC to succeed. These conditions are expressed as relations between the adiabatic ramp time, the low-energy spectrum of our model, and the rate of change of ground-state. In section\ \ref{sec3aq} we compare the performance of a linear adiabatic passage for an Ising model and for a spin-boson model, both of which implement frustrated antiferromagnetic interactions. Section\ \ref{sec3db} explains the problems in such comparison, such as the differences in the spectra of both models, instantaneous correlations, etc. Section\ \ref{sec3db} shows how to build a fair comparison between mediated and direct interactions without any dependency on the parameterization. The results of this fair comparison are shown in section\ \ref{sec3nr}, where we find evidence for the actual role of quantum fluctuations in bosonic mediated interactions. Section \ref{Sec:4} summarizes our findings and the outlook for future work based on them.

\section{The spin-boson quantum simulator}\label{Sec:2}

\begin{figure}[t!]
\begin{centering}
\includegraphics[width=1.\columnwidth]{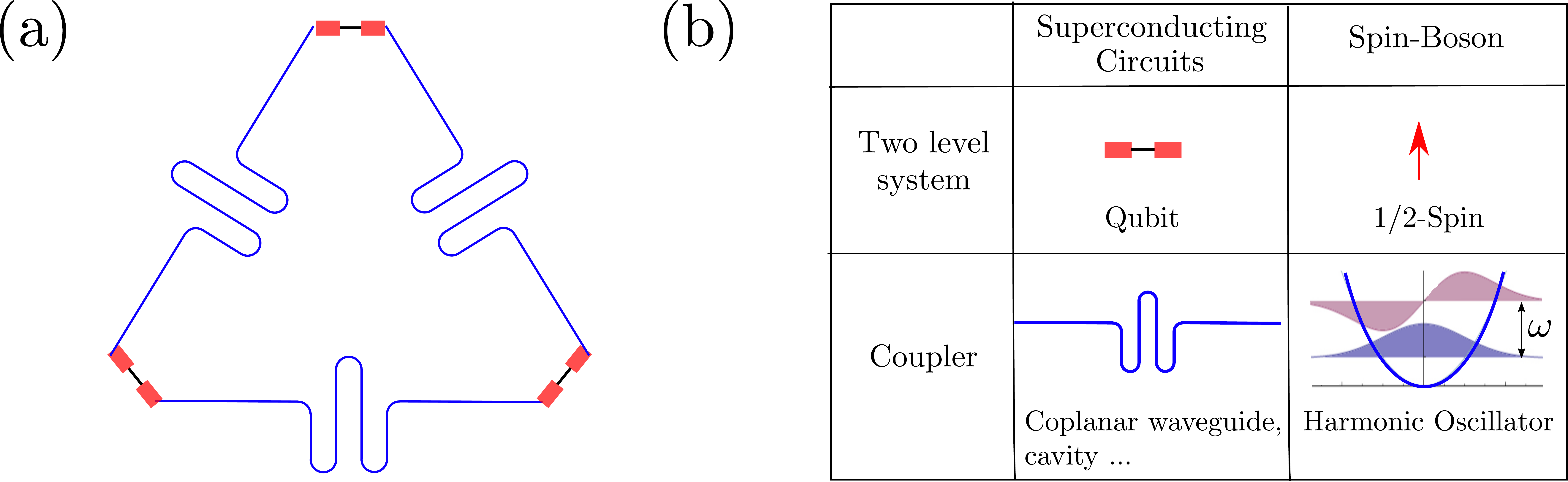}
\par\end{centering}
\vspace{0.2cm}
\caption{(a) Sketch of three superconducting qubits coupled to inter-spaced resonators via superconducting waveguides. (b) Table for the equivalence of each element in the superconducting circuit and in the spin-boson model.
}\label{Fig1}
\end{figure}

Let us take a general quantum Ising Hamiltonian
\begin{equation}
    H = \sum_{ij} J_{ij} \sigma^x_i \sigma^x_j + \sum_i (B_i \sigma^x_i + h_i \sigma^z_i).
    \label{eq:Ising}
\end{equation}
In absence of transverse magnetic field $h_i=0$, this Hamiltonian can represent many different NP-hard logical and combinatorial optimization problems. In those mappings the problem is defined as a non-negative Boolean function whose solution satisfies $f(s_1,\ldots,s_N)=0$, with $s_i\in\{0,1\}$. This function is re-expressed as the Ising model equation\ \eqref{eq:Ising} with suitable constants $J_{ij}$ and longitudinal fields $B_i$, replacing the binary numbers $s_i$ with Pauli matrices $\sigma^x_i$ that have two possible eigenstates $\ket{0}$ and $\ket{1}$. Each of the solution of the problem $f(s_1,\ldots,s_N)=0$ is associated with a ground state in a product form $\ket{s_1,s_2,\ldots,s_N}$ ---a classical state with well defined spin orientations that gives an answer to the problem.

The goal of the quantum annealer is to prepare those ground states, using a strategy of annealing or ``tempering", where the quantum fluctuations introduced by $h$ are initially large and are progressively decreased, until $h=0$ and the desired state is obtained. The experimental implementation of a quantum annealer requires a set of qubits with a large connectivity and tunable interactions, that allows embedding any effective $J_{ij}$. Unfortunately, qubit-qubit interactions are usually weak --as in the case of magnetic dipoles in atomic qubits--, they are fixed --e.g. two flux qubits placed close to one another couple via an inductance that is fixed by the geometry--. For this reason, spin models such as the one in equation (\ref{eq:Ising}) are obtained as the effective low-energy dynamics of analogue quantum simulators with mediated and flexible interactions.

In this work we introduce interactions that are mediated by bosonic degrees of freedom. This is the case of trapped ions, where atomic degrees of freedom interact via the phonons of the ion crystal, creating an effective spin model\ \cite{porras2004effective}. It is also the case of superconducting circuits, where a transmission line or a cavity can couple distant superconducting qubits\ \cite{majer2007coupling,sillanpaa2007coherent,niemczyk2010circuit,forn2017ultrastrong}. We describe this paradigm of mediated interactions using a spin-boson Hamiltonian, where the spins are our qubits and the harmonic oscillators are our bosonic mediators. We have in mind a topology such as the superconducting architecture in figure\ \ref{Fig1}, where there are $N$ spins and $N_b$ bosonic degrees of freedom, each of which connects to two or more spins:
\begin{align}\label{Eq:SBQA}
 H_\text{SB}= \sum_{i=1}^N \sum_{r=1}^{N_b}\sigma_i^x \left(g_{ir} b_r^\dagger +g_{ir}^* b_r\right) +  
 B_i \sigma_i^x  +   \frac{\omega_0}{2} \sum_{i=1}^{N}\sigma_i^z + 
 \sum_{i=1}^{N_b} \omega_r  b_r^\dagger b_r.
\end{align}
The Pauli matrices $\sigma_i^\alpha$ with $\alpha\in\{x,y,z\}$ represent the qubit degrees of freedom. Each qubit couples to one or more bosonic operator, with Fock operators $\{b_r,b_r^\dagger\}$ and coupling strength $g_{ir}$. In addition to this, we allow for a magnetic field in the computational basis $B_i$ and we consider uniform gaps $\omega_0$ in the qubits. The  $\omega_r$ are the frequencies of the oscillators. We will typically take $\omega_0=1$ as unit of energy in the rest of the manuscript.

The model equation \eqref{Eq:SBQA} encodes an effective spin-spin interaction that is revealed by numerical simulations at low energies, and also by a polaron transformation\ \cite{kurcz2014hybrid} that entangles spins and oscillators $U=\exp\left[-\sum_{r,i}\sigma^x_i  (g_{ir} b_r^\dagger +g^*_{ir}b_r )/\omega_r\right].$ This transformations creates the effective model $H_\text{pol}=U^\dagger H_\text{SB} U$
\begin{align}\label{Eq:sb}
 H_\text{pol} = \sum_{i,j=1}^{N} J_{ij} \sigma_i^x \sigma_j^x + \sum_{i=1}^N h_i \sigma_i^x  
 + \frac{\omega_0}{2}\sum_{i=1}^N  e^{-2\sum_r|\frac{g_{ir}}{\omega_r}|^2 } \sigma_i^z \theta^\dagger_i(-\sigma_i^x) \theta_i(\sigma_i^x) + \sum_{r=1}^{N_b} \omega_r b_r^\dagger b_r .
\end{align}
where $\theta_i(x) = \exp(2x\sum_r g^*_{ir}b_r/\omega_r)$. This model has the coupling $J_{ij}=- \sum_r \mathrm{Re}(g_{ir}g^*_{jr}/\omega)$ and magnetic field $h_i$ required to encode useful problems, as in Eq.\  \eqref{eq:Ising}. We will focus on frustrated antiferromagnetic interactions $J_{ij}\geq 0$, because typical hard problems belong to this family\ \cite{farhi2001quantum}. Our goal is to prepare the ground states of Hamiltonian in equation \eqref{Eq:SBQA}, revealing the magnetic order dictated by the effective model in equation \eqref{Eq:sb} and studying how this task can be different in this model than in the original Ising Hamiltonian.

In order for quantum annealing to work and prepare a good ground state, it needs a Hamiltonian that has quantum fluctuations: i.e. elements that induce quantum tunneling between states in the computational basis, allowing a flow of probability towards the ground state. The polaron transformation shows that there are two sources of quantum fluctuations in the spin-boson model equation \eqref{Eq:SBQA}: the transverse magnetic field $\sigma^z$ and the fluctuations induced by the occupation of the bosonic modes, due to the polaron term $\theta_i(\sigma^x_i) = \exp(2\sigma^x_i\sum_r g^*_{ir}b_r/\omega_r)$. When the resonators are highly detuned from the qubits $\omega\gg \omega_0$, the bosonic fluctuations cancel, $\theta\to 1$, and we recover a pure Ising interaction. This limit can be explained via a Schrieffer-Wolff transformation \cite{schrieffer1966relation,kurcz2014hybrid} and it is the usual one in the dispersive coupling of trapped ions\ \cite{porras2004effective} or superconducting qubits. However, we wish to study the quantum annealing in the full model, revealing the positive or negative effect of bosonic quantum fluctuations in the setup.

\section{Adiabatic quantum computer with mediated interactions}\label{Sec:3}

In this section we will study the quantum annealing of an effective spin model generated by a tuneable spin-boson quantum simulator, equation\ \eqref{Eq:SBQA}. Our study is structured in three sections. First, in section\ \ref{sec3ac} we will discuss how to prepare the ground state of a Hamiltonian by interpolating the parameters in the model, from a simple to prepare state, to the actual model we wish to solve. We will relate the efficiency of the process to the tuning of a single parameter, showing how the rate of change of this control is imposed by the instantaneous gap of the problem. In section\ \ref{sec3aq} we will make a first attempt at a comparison between an adiabatic passage on the spin-boson model equation\ \eqref{Eq:SBQA} and the full Ising Hamiltonian equation\ \eqref{eq:Ising}. This comparison will reveal that a linear ramp of the interactions in the spin-boson model performs better than a linear ramp in the Ising model. We will show in section\ \ref{sec3db} that this discrepancy is due to a mismatch between the effective spin-boson and Ising models, so a better comparison requires to engineer passages whose constants depend in a nonlinear way with the tuning parameter, which is done in section\ \ref{sec3cp}. Numerical results with those passages, presented in section\ \ref{sec3nr}, show that some bosonic modes interchange population with spin errors and affect the results of quantum annealing.


\subsection{Adiabaticity condition}
\label{sec3ac}

The Quantum Adiabatic Algorithm aims to find the ground-state of a target Hamiltonian via adiabatic evolution from an easy-to-prepare initial state.  Specifically, one uses a system of qubits with initial Hamiltonian $\Delta H$ and target Hamiltonian $H_0$. The passage to interpolate between those Hamiltonians can be written as:
\begin{align}\label{Eq:QA}
 H(s)= s H_0 + \left(1-s\right) \Delta H.
\end{align}
The parameter $s$ has to be slowly varied from $0$ to $1$ during evolution. How slowly depends on the scaling of the minimum gap of the Hamiltonian during the passage, as we will see. Adiabatic quantum algorithm provides a new approach to tackle several computational problems, as NP-hard, for which the performance of classical computers seems to be poor\ \cite{farhi2001quantum}.

The performance of  AQC is explained by any of the adiabatic theorem variants\cite{de2010adiabatic}. We label the $n$ eigenstates and eigenvectors along a passage as  $\ket{\psi_i(s)}$ and $E_i(s)$ with $i=0,\dots N$. A simple version of local adiabatic condition imposes \cite{amin2009consistency}:
\begin{align}\label{Eq:AT}
\left| \frac{\hbar\langle \psi_1(t)|\partial_t|\psi_0(t) \rangle}{\Delta(t)} \right|  = \frac{\left|\langle \psi_1(s)|(\partial_{s}H)|\psi_0(s) \rangle \right|}{T} \frac{\hbar}{\Delta(s)^2}\ll1
\end{align}
at all times $t$ along the passage. This equation focuses on suppressing non-adiabatic transitions from the ground-state to the first excited state, using the gap $\Delta=E_1-E_0$ as reference. The condition assumes that the Hamiltonian parameter $s$ changes as a function of time $s=s(t/T)$, where $T$ is total time of evolution and $s(0)=1$ and $s(1)=0$. A more rigorous versions of adiabatic theorem can be found in \cite{jansen2007jansen}.

The success of the adiabatic quantum computer is ensured for a total time of evolution $T$ that scales as the minimum gap squared $\Delta_m^{-2}$. The hardness of a problem is thus determined by how the minimum gap grows with the system size $\Delta_m(N)$. For example, a problem would be solvable if the minimum gap along the passage occurs at a second order phase transition, as those critical points have a polynomial law $\Delta\sim N^{-z}$ with $z$ the dynamical critical exponent\ \cite{Polkovnikov2005universal}. Unfortunately, the situation is not so optimist for NP-hard optimization problem, where the minimum gap seems to become exponentially small with increasing number of bits\ \cite{young2008size, altshuler2010anderson, knysh2016zero}. 

In this work we will not discuss the scaling of the minimum gap for the spin-boson model as this is a very difficult question, which is problem-dependent ---and subject to controversy already in the pure spin problem\ \cite{laumann2012quantum,knysh2010relevance, choi2011different,altshuler2010anderson,knysh2016zero}---. Instead, we will assume a plausible situation: that the gap scales similarly for the spin-boson model and for the Ising model to which it is related, with differences in the constant factors of this law. This is suggested by earlier simulations\ \cite{kurcz2014hybrid}, and it allows us to focus on what part of the performance differences originate in the mediated interactions and the fluctuations of the bosonic particles.

\subsection{AQC for a frustrated antiferromagnet}\label{sec3aq}

We compare the performance of the adiabatic quantum algorithm for direct and mediated interactions. We choose a spin model with nearest-neighbors antiferromagnetic interactions
\begin{align} \label{Eq:H_spins}
H_0=\sum_{i=1}^{N} \sigma_i^{x}\sigma_{i+1}^{x} 
\end{align}
with periodic boundary conditions $\sigma_{N+1}=\sigma_N$ and odd number of spins. This configuration produces a frustrated ground-state with $2N$-fold degeneracy. This model is reproduced at low energies by a Hamiltonian \eqref{Eq:SBQA} with equal number of spins and bosons $N_b = N,$ with alternating couplings $g_{i,i+1}=-g_{i+1,i+1}=\sqrt{\omega}$ (cf. see figure\ref{Fig1}). We will work with exact diagonalizations in small systems, due to the cost of including the bosonic degrees of freedom in the simulation.

\begin{figure}[t!]
\begin{centering}
\includegraphics[width=1.\columnwidth]{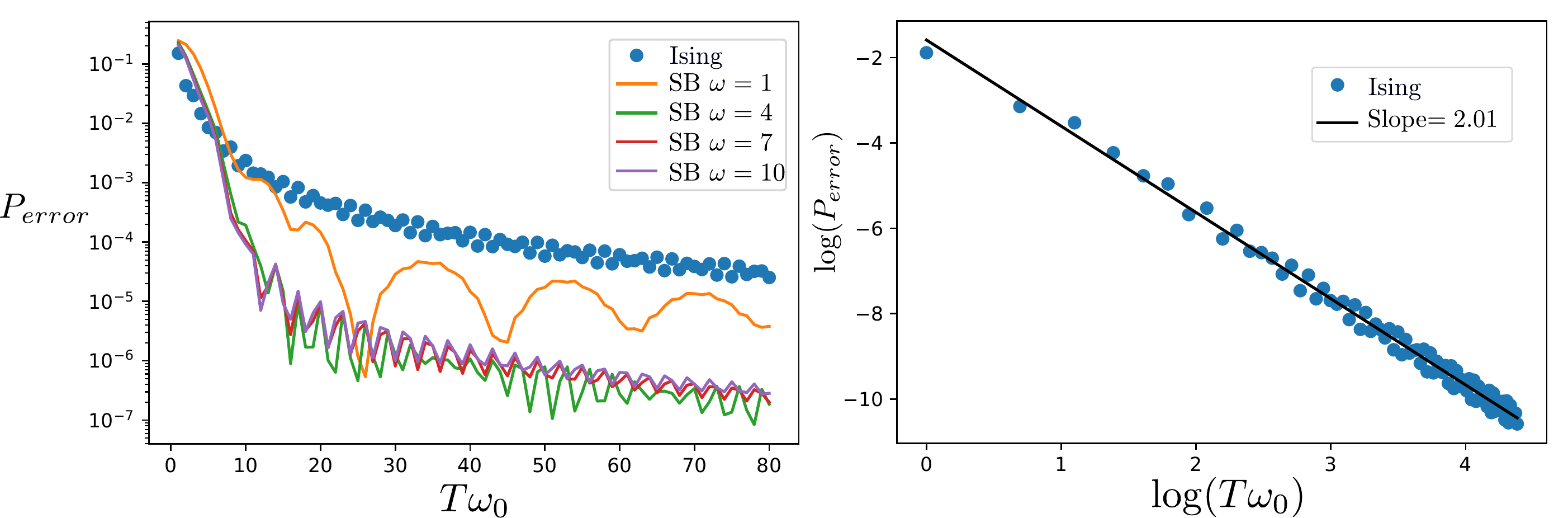}
\par\end{centering}
\caption{Left: probability of error after a passage of total time $T$ for an antiferromagnetic target Hamiltonian given in equation\ \eqref{Eq:H_spins}, with $N=3$ spins.  Dots corresponds to Ising passage, equation\ \eqref{Eq:lin_passIs},  while solid lines corresponds to Spin-Boson, equation\ \eqref{Eq:lin_passSB}, for several frequencies. Right: probbility of error as a function of $T$ for Ising passage in log-log scale. The solid line is a fit to a straight line with slope equal to $2$. So the Ising case is represented by a law
$P_{error}\sim T^{-2}$.
}\label{Fig2}
\end{figure}

In our first attempt we will compare linear adiabatic passages for both the Ising and spin-boson model:
\begin{align}
 H_{\rm I}(s) =& s \sum_{i=1}^{N} \sigma_i^{x}\sigma_{i+1}^{x} + \frac{\left(1-s\right)}{2} \sum_{i}   \sigma_i^z  \label{Eq:lin_passIs} \\
 H_{\rm SB}(s)=& s\sqrt{\omega} \sum_{i=1}^N \sigma_i^x \left( b_i - b_{i+1}+ {\rm H.c}\right) +  \frac{\left(1-s\right) }{2}\sum_{i=1}^{N}\sigma_i^z + 
  \omega\sum_{i=1}^{N}   b_r^\dagger b_r. \label{Eq:lin_passSB}
\end{align}
In both models we use the same parametrization $s = t/T$, with a time $t\in[0,T]$.

Figure\ \ref{Fig2} shows the probability of error $P_{error}$ in the preparation of the ground state of $H(1)=H_0$, as a function of the total time $T$ for $N=3$ spins. At long times $P_{error}$ decays algebraically with the total time, instead of decreasing exponentially in $T$, as one would expect from a Landau-Zener theory\ \cite{landau1932theorie}. An interpretation is that our passage behaves as ``half" a Landau-Zener transition: from a ferromagnet pointing down in the Z axis to a nontrivial state with magnetization pointing along the X direction. The full Landau-Zener would be recovered if the Hamiltonian was further changed to end up in a ferromagnet along the z-axis but in opposite direction. The ``half" Landau-Zener treated in Ref.\ \cite{de2010adiabatic} has an error $P_{error}\sim T^{-2}$, which coincides with the results for the Ising passage as can be seen in the panel at right of figure \ref{Fig2}.

Another feature in our simulations is that the probability of error depends on the nature of the coupling. The left panel of figure\ \ref{Fig2} shows that the error of the spin-boson model at any frequency is smaller than the one for the Ising model by what seems a constant prefactor, once we pass the diabatic limit and the frequency of the resonator is deep in the perturbative limit $\omega\sim 4-10$. The model with $\omega=1$ exhibits larger oscillations with slower period, but still stays below the Ising Hamiltonian. We will discuss the origin of all these differences and of the oscillations in the following sections.

\subsection{Discrepancies between lineal ramps for Ising and spin-boson}\label{sec3db}

Figure\ \ref{Fig2} does not provide a fair comparison between spin-boson and Ising models. The reason is that the effective coupling and transverse field along the passage for spin-boson and Ising are quite different. This can be seen via the polaron transformation described in section\ \ref{Sec:2}. This transformation predicts an effective nearest-neighbors coupling and transverse field given by:
\begin{align}
 J^{\rm SB}(s)& = s^2 \label{Eq:rt1}\\
 h^{\rm SB}(s) &=  \frac{1-s}{2}e^{-2s^2}  \theta^\dagger_i(-\sigma_i^x) \theta_i(\sigma_i^x)\label{Eq:rt2}
\end{align}
with operators $\theta_i(x) =\exp\left[{-2x\sum{\sqrt{\frac{s}{\omega}}(b_i-b_{i+1})}}\right]$ introduced in section \ref{Sec:2}.  This functional dependence is different from the one in the Ising passage where $J^{\rm I}= s$ and $h^{I}= 1-s$ and this occurs even in the perturbative regime, $\omega\gg1$.

\begin{figure}[t!]
\begin{centering}
\includegraphics[width=1.01\columnwidth]{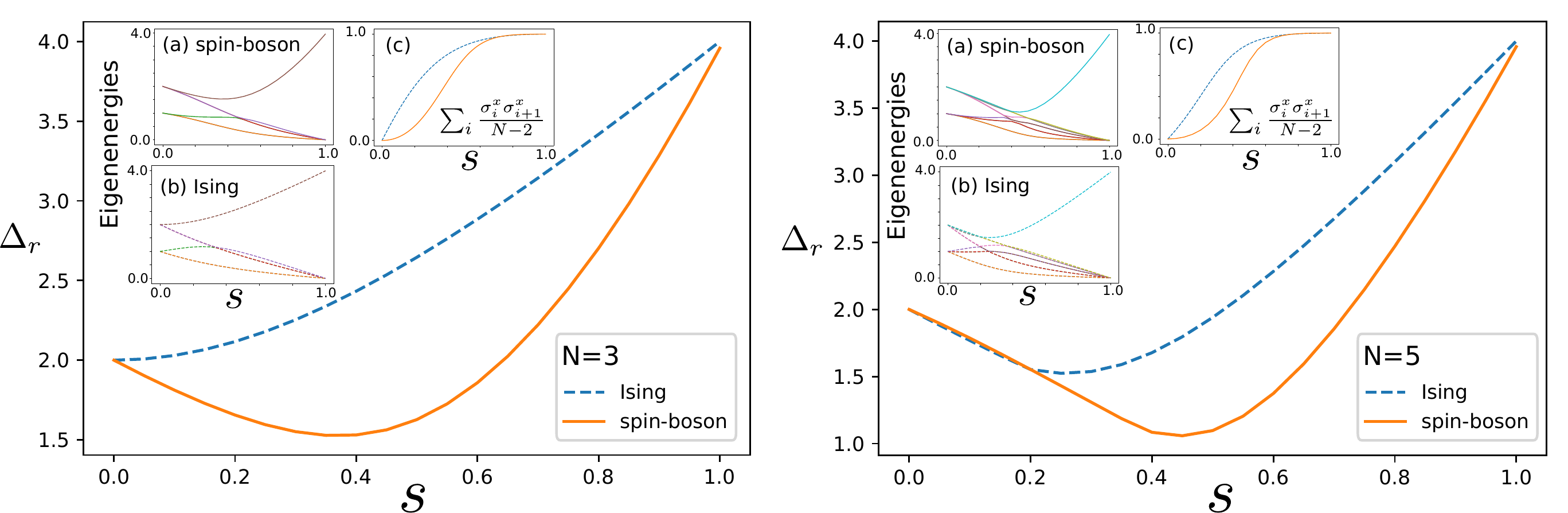}
\par\end{centering}
\caption{Relevant gap as a function of the parameter $s$ along passages defined for Ising in equation\ \eqref{Eq:lin_passIs}, dashed line, and  for spin-boson in equation\ \eqref{Eq:lin_passSB}, solid lines. The relevant gap is $\Delta_r =E_{2N}-E_0$, which is the difference between excited state $2N$ and ground-state,  being $N$ the number of spins. The data for spin-boson case corresponds to a resonator frequency $\omega = 4$. Left panel corresponds to size $N=3$ and right one to $N=5$. The insets in each panel show:  low-energy spectrum of eigenstates $E_1-E_0$, $E_2-E_0$, $\dots$, $E_{2N}-E_0$ for spin-boson (a) and  Ising (b). The inset (c) corresponds to the ground-state expected value of the operator $O = \sum_{i=1}^N\sigma^x_i\sigma^x_{i+1}/(N-2)$, where $\sigma^x_{N+1}=\sigma^x_1$.
}\label{Fig3}
\end{figure}

The discrepancy in constants between ramps has three consequences. First, it implies that the rate of change of ground-state in equation\ \eqref{Eq:AT}, $\left|\langle \psi_1(t)|\partial_{t}|\psi_0(t) \rangle \right|$, is different for direct and mediated interactions, as it depends on the ratio $J/h$. Second, it produces differences in the gap along each of the passages as the gap depends on the absolute value of $J$ and $h$ (not only on the ratio). Finally, the quantum fluctuations $\theta(\sigma^x)$ also affect the total gap of the spin-boson, creating a complicated situation that needs to be clarified.

This discrepancy affects the original goal of this paper, which is to understand how the dynamics of bosonic fluctuations affect quantum annealers, assuming all things equal ---in particular the gap---. The main tool to correct these problems will be to adjust the spin-boson and Ising spin models so that they have the same low-energy sector, creating states that have the same correlations and the same energy levels all along the passage. This will be done in section\ \ref{sec3cp}, by reparameterizing the spin-boson model and multiplying the Ising model by an overall constant that ensures both have the same norm in the spin sector. However, in order to properly define this change, we must first compare carefully the eigenspaces of the full spin-boson model and of the Ising model. In particular, we identify eigenstates that have similar properties in the spin sector and find the ``relevant" gap between the subspace of ground states and the first excited state that has an error with respect to the desired configurations of our problem Hamiltonian $H_0.$

As shown in insets (a) and (b) of figure\ \ref{Fig3}, the spectrum for direct and mediated passages contains a set of $2N$ low-energy eigenstates which become degenerate with the ground-state at the end of the passage. This degeneracy is due to the frustrated nature of the target Hamiltonian. A finite population in any of those degenerate states will not imply an error at the end of the passage. Thus, the relevant processes that produce errors are those in which the evolution brings the system to the first state outside of the ground-state degenerate manifold. We consider the ``relevant" gap, denoted by $\Delta_r$, to be the energy between ground-state and that state, which for our frustrated Hamiltonian is the $2N$-th excited state (ground-state is labeled by index $0$).

The relevant gaps along the passage for mediated (solid lines) and direct coupling (dashed lines) are represented in figure\ \ref{Fig3} as a function of dimensionless parameter $s$ for resonator frequency $\omega=4$. The left panel corresponds to $N=3$ spins and right one to $N=5$. The minimum gap for spin-boson is smaller than in Ising passages in both cases, and also for other values of resonators frequencies. However, the rate of change of the ground state is also slower in the spin-boson model. Indeed, as show in the magnetization plots, the spin-boson model abandons the paramagnetic phase more slowly than the Ising Hamiltonian. As a result, following the qualitative model in section\ \ref{sec3ac}, the spin-boson model results more adiabatic and has better errors (cf. figure\ \ref{Fig2}a).


\begin{figure}[t!]
\begin{centering}
\includegraphics[width=0.9\columnwidth]{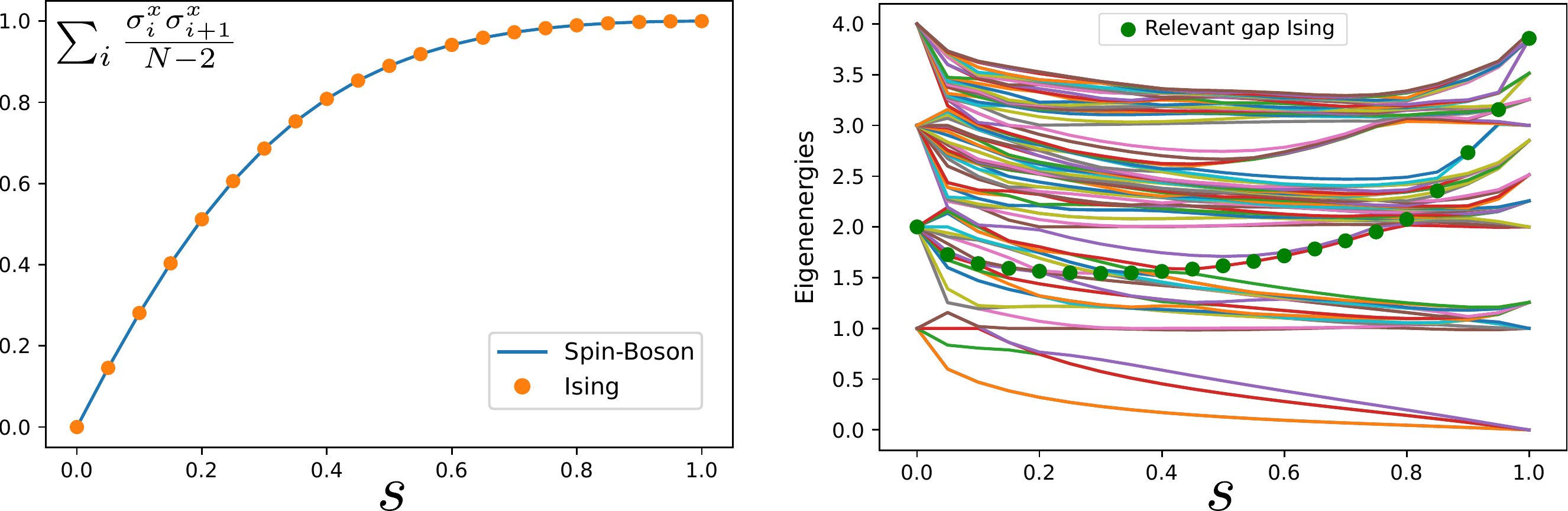}
\par\end{centering}
\caption{Ground-state correlations and energy levels of the fair parameterizations equations\ \eqref{Eq:repSB} and \eqref{Eq:repIS} for frequency $\omega=1$ and size $N=3$. These parameterizations were designed so that spin correlations in the ground-state and the relevant gap are the same at each step of the passage $s$. Left: the ground-state expected value $O= \sum_{i=1}^N \sigma^x_i\sigma^x_{i+1}/(N-2)$ for spin-boson (solid) and Ising (bullets). Right: eigenenergies of the spin-boson passage appears as solid lines. The circles represents the relevant gap for the Ising passage. This relevant gap is defined as the gap between ground-state and the first excited state that escapes from the degenerate ground-state manifold at $s=1$.}
\label{Fig4}
\end{figure}  

\subsection{Correct parameterizations}\label{sec3cp}

We are now going to reparameterize both models to implement the same passage in the spin sector, eliminating the differences that were shown in figure\ \ref{Fig3}. The new passages for spin-boson and Ising will meet two conditions. First, the rate of change of eigenstates have to be the same for comparable spin sectors. Second, we need to scale up (or down) the norm of the Ising model so that both Hamiltonians have the same instantaneous energy ---and the same relevant gap--- along the whole passage.

First condition can be met by using parameterization $s=s(\lambda)$ and $\lambda=t/T$, so that $\langle \psi_1(s \circ \lambda)|\partial_t |\psi_{gs}(s \circ \lambda)\rangle$ has similar values for spin-boson and Ising models. To do so, we use a parameterization for direct coupling $s_{\rm I}(\lambda)=\lambda$  and parameterization for spin-boson passage $s_{\rm sb}(\lambda) = O_\text{SB}^{-1}(O_I(\lambda))$. The $O_\text{SB}$ and $O_{I}$ are the expected value of operator $ O = \sum_{i=1}^N \sigma^x_i\sigma^x_{i+1}/(N-2)$, along the old, $s(\lambda) =\lambda$, parameterization for spin-boson and Ising models, respectively.  We notice that operator $O$ defines unequivocally the ground-state along spin-boson and Ising passages. In insets (c) of figure\ \ref{Fig3}, the ground-state expected value of $O$ is computed along passages equations\ (\ref{Eq:lin_passIs}) and (\ref{Eq:lin_passSB}).

The second condition is implemented by making ``relevant" gaps in spin-boson and Ising to be the same, multiplying the Ising Hamiltonian by a function $c(\lambda)$ at each point of the passage. This function is $c(\lambda)=\Delta^{\rm SB}_r(\lambda)/\Delta^{\rm I}_r(\lambda)$, where $\Delta^{\rm SB(I)}_r(\lambda)$ is the relevant gap along the new parameterization for spin-boson (Ising). 

First criteria can be numerically implemented by computing the inverse functions of $O_\text{SB}$ and $O_{I}$ for the old parameterization. The implementation of second criteria is more subtle. In fact, it is difficult to determine the target excited state corresponding to the "relevant" gap for spin-boson when frequency of resonators is small $\omega \sim 1$. The difficulties are due to eigenstates with different bosonic occupations lying in  a narrow interval of energies. What we do is to define the target eigenvalue for the spin-boson model such that it maximizes: $$ p_i = \langle\psi^{I}_r | \rho_i^{sb} | \psi^{I}_r\rangle $$ where $ \rho_i^{sb} = tr_{b}{| \psi_i^{sb} \rangle \langle \psi_i^{sb} |}$ is the density matrix of spin-boson eigenstates $i$ where the bosonic degrees of freedom have been traced out.  The state $| \psi^{I}_r\rangle$ is the target state for direct coupling which we know is the $2N$ excited state. Thus, passages for spin-boson and Ising  Hamiltonians are:
\begin{align}
 \mathcal{H}_{\rm SB}(\lambda) &= H_{\rm SB}\left( s_{\rm SB}(\lambda)\right)\label{Eq:repSB}\\
 \mathcal{H}_{\rm I} (\lambda) &= c(\lambda) H_{\rm I}\left(s_{\rm I}(\lambda)\right)\label{Eq:repIS}
\end{align}
The expected value of operator $O$ and the spectrum of energies is represented in figure \ref{Fig4} for both of these passages. It can be seen that correlations are the same along the passages (left panel). The relevant gap for Ising model appears as circles at right panel. Our method allows the determination of the relevant gap even at small $\omega$, where there are many eigenenergies in the spin-boson model.

\subsection{Numerical results}\label{sec3nr}

\begin{figure}[t!]
\begin{centering}
\includegraphics[width=1.\columnwidth]{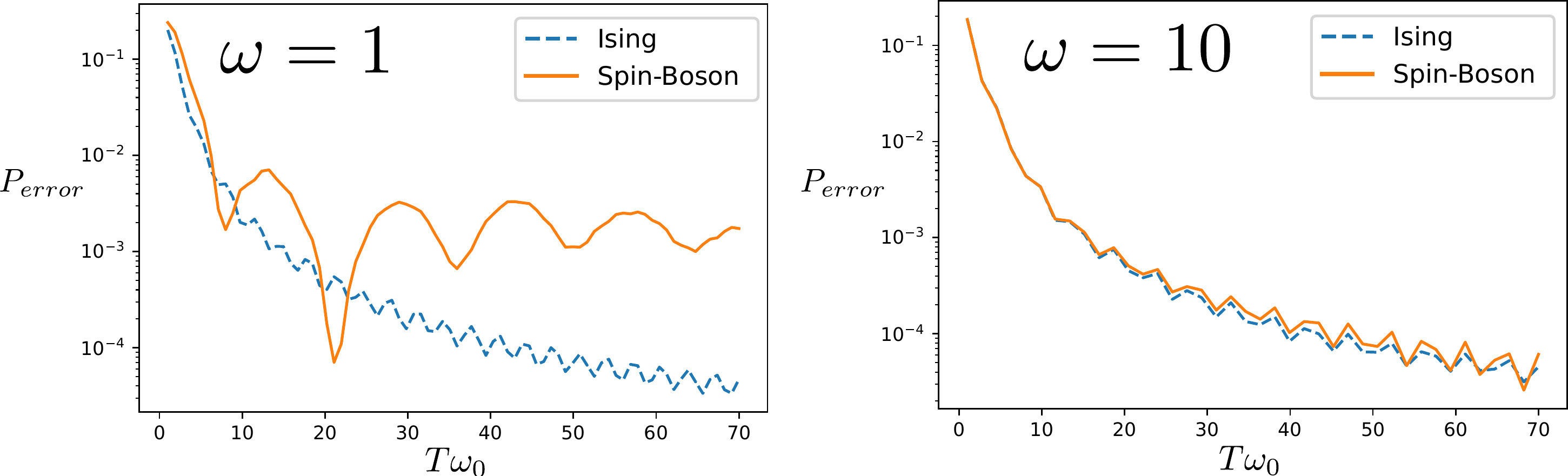}
\par\end{centering}
\caption{ Probability of error, $P_{error}$, at the end of a passage of total time $T$. Each of the panel corresponds to a value of resonator's frequency, left $\omega=1$, and right $\omega = 10$. The passages for spin-boson and Ising Hamiltonian use parameterization equations\ \eqref{Eq:repSB} and \eqref{Eq:repIS}, respectively.
}\label{Fig5}
\end{figure}

The rescaling and reparameterization above enable a fair comparison of quantum annealing in spin-boson and Ising models. We now present the results of numerical simulation using those passages. The probability of error after a passage of total time $T$ appears in figure\ \ref{Fig5} for direct (dashed) and mediated interactions (solid), using resonator frequencies $\omega = 1$ (left panel) and $\omega = 10$ (right panel). Note how both models coincide in the perturbative regime $\omega=10$, in which quantum fluctuations due to the bosonic modes vanish. This means that the comparison with the parameterization equations\ \eqref{Eq:repSB} and \eqref{Eq:repIS} only reveals differences caused by bosonic fluctuations, as desired.

The main findings of this paper can be extracted from the panel corresponding to $\omega=1$ in figure \ref{Fig5}. They are: (i) the dynamics of bosons affects weakly the performance of a quantum annealer for relatively fast passages (times $T\omega_0< 10$), (ii) the presence of bosonic fluctuations limits the accuracy of annealing at large times, $T\omega_0\gg 1$ and (iii) there are finite values of total evolution time at which bosonic fluctuations improve significantly the annealing (for example at $T\omega_0\sim 20 $). We have checked that these conclusions hold for other values far from the perturbative regime.

\begin{figure}[t!]
\begin{centering}
\includegraphics[width=1.\columnwidth]{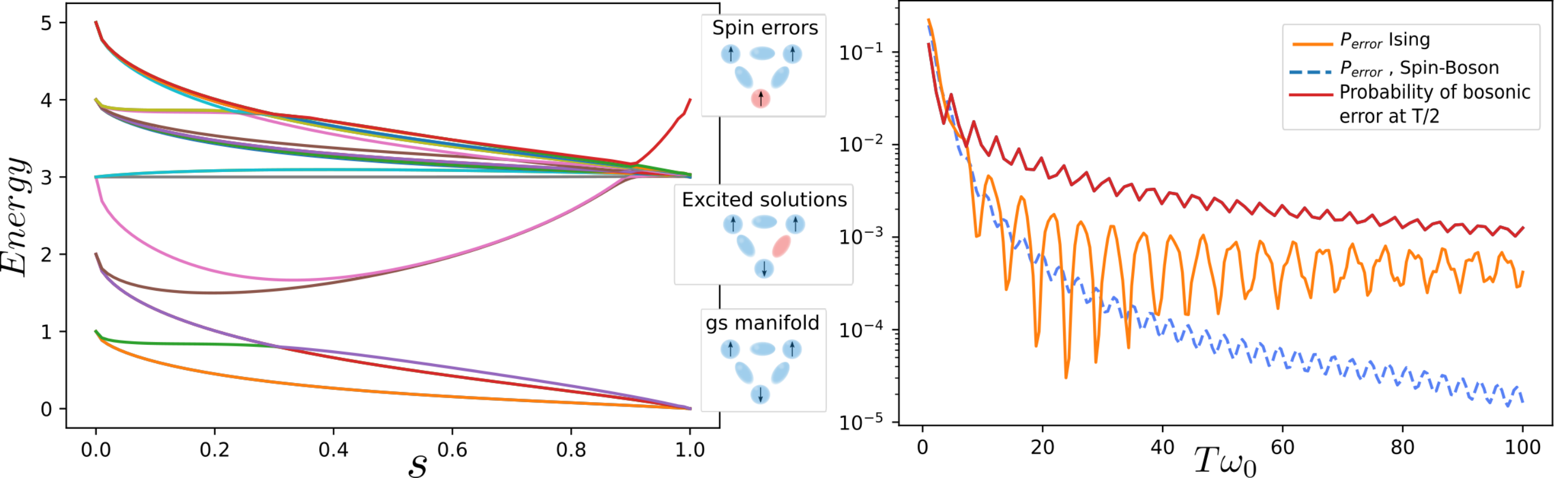}
\par\end{centering}
\caption{Left plot: low-energy spectrum for spin-boson passage equation\ \eqref{Eq:repSB} for $\omega=3$ and size $N=3$. We can divide the eigenstates in three groups which depend on the value of energy at $s=1$. The spin sector of the ground-state manifold is a frustrated antiferromagnet. The group of excited solution are states with the same spin correlations as the ground-state manifold but with larger bosonic occupation. The group of spin errors contains a ferromagnetic ground-state, so one spin is flipped respect to ground-state manifold. Note how an avoided level crossing between excited solutions and spin error occurs at $s\sim 0.9$. Right plot: quantum annealing error $P_{error}$ vs. total time of evolution $T$ for passages\ (\ref{Eq:repSB}) and (\ref{Eq:repIS}). The case for spin-boson appears as a solid orange line and the one for Ising as a dashed blue line. The red solid line represents the probability to have an excited solution at the middle of the passage, that is, at time of evolution $t=T/2$. 
}\label{Fig6}
\end{figure}

These results are explained by the spin-boson instantaneous eigenstates. In figure\ \ref{Fig6} we show the spectrum of $\omega=3$, which is less crowded than the one of $\omega =1$ but has the same features. The low-energy eigenstates include \textit{spin errors} ---states with spin flips--- and \textit{excited solutions} ---which have the same state in the spin sector as the final adiabatic solution, but with some extra bosons.

As shown by the left panel of figure\ \ref{Fig6}, the groups of spin errors and excited solutions cross at approximately $s\approx 0.9.$ In general, this implies the existence of a new time scale $T^\star$ settled by the splitting at this level crossing. When the passage is adiabatic with respect to this crossing $T\ll T^\star$, there can be a transfer of population between excited solution and spin errors. In addition to this, the excited solutions can have some residual population already at the beginning of the passage, which is more or less preserved right until the crossing point [cf. right panel in figure\  \ref{Fig6}].

This implies a complex picture in which the performance of the quantum annealer depends radically on how the bosonic and spin errors mix during evolution. We can understand the total errors as a coherent sum of two contributions: (i) nonadiabatic excitations from the ground state into a spin error and (ii) spin errors that originate from excited solutions. Out of these phenomena, (ii) is only present in the spin-boson model.

For short times, $T\ll T^*$ or $T\omega_0< 10$, the population of the excited solutions is preserved and the Ising and spin-boson models behave identically. When we slow down the passage into a fully adiabatic limit $T\gg T^\star$ $(T\omega_0 \gg 90)$, the final error is dominated by contributions from initial excited solutions that are converted into spin errors. Finally, there is an intermediate regime $T\sim T^\star$ ($T\omega_0\sim 10-60$), where both processes interfere and there can be a net transfer of population from spin errors into excited solutions, leading to an improvement of the overall performance at finite times.

In summary, we have shown evidence that the quantum fluctuations that mediate interactions can both improve or worsen the adiabatic solutions, by extracting or injecting errors into the spin sector. In the particular adiabatic passage that we have constructed the injection of errors dominates at long times. However, this work suggests other passages where the frequency of bosons is adjusted in time, so as to minimize the initial contribution of excited solutions and maximize the cooling or error correction induced by the bosons.

\section{Conclussions}\label{Sec:4}

This work has studied quantum annealing for devices where bosonic degrees of freedom mediate an interaction between spins or qubits.
Our analysis has been general enough to cover a wide range of bosons frequencies, from the perturbative regime $\omega\gg 1$ to the strong coupling regime $\omega\sim 1$ (units of qubit frequency). Our goal was to compare the direct and mediated case in a fair way, and to understand how the auxiliary modes affect quantum annealing. This comparison has been done for a toy model composed of a few qubits in which the target state is a frustrated antiferromagnet.

We have seen that, apart from spin errors (spin flips), there are what we called excited solutions: states with the same spin correlations as the ground-state but with larger bosonic occupations. When the frequency of the resonators $\omega$ is small, the excited solutions cross the space of spin errors. The splitting of those level crossings determines a time scale $T^\star$, such that for $T\gg T^\star$, there is a transfer of population from spin errors to excited solutions and vice versa ---quantum annealing can therefore improve or worsen depending on the direction of this population transfer. We have found that this self-correction phenomenon improves the outcome of quantum annealing for intermediate times $T\sim T^\star$.

We believe that this picture holds for quantum annealing with mediated interactions in more complex systems than the one used in here. The polaron transformation and second-order perturbation theory reveal the existence of general excited solutions and of avoided level crossings between those states and spin flips, provided the resonator frequency is low enough.

In summary, we have settled the bases to study quantum annealing for mediated interactions and we have given arguments to support the utility of such a study, especially in the strong coupling regime. Our results suggest further investigation of other adiabatic passages where we make better use of the excited solutions by controlling both the coupling strength $g$ as well as the resonator frequency $\omega$. 

Finally, we have not addressed an important and difficult question: how bosonic fluctuations renormalize the gap and how this may influence the overall efficiency of quantum annealing. The answer is very likely to be problem dependant and the scaling of the gap in interesting, NP-hard problems, is a pretty difficult question that has not been fully clarified yet for the simpler case of direct spin interactions\ \cite{altshuler2010anderson, knysh2010relevance, knysh2016zero, laumann2012quantum}. We plan to study this question in the context of boson-mediated interactions in a future work.

\section*{Acknowledgments}
This work was supported by MINECO/FEDER Project FIS2015-70856-P, CAM PRICYT Project QUITEMAD+ S2013/ICE-2801.
M. P. acknowledge support from Juan de la Cierva IJCI-2015-23260 and Proyecto de la Fundacion Seneca 19907/GERM/15.

 
\bibliography{./spin_boson}

\end{document}